\documentclass[10pt,conference]{IEEEtran}

\makeatletter
\def\ps@headings{%
\def\@oddhead{\mbox{}\scriptsize\rightmark \hfil \thepage}%
\def\@evenhead{\scriptsize\thepage \hfil \leftmark\mbox{}}%
\def\@oddfoot{}%
\def\@evenfoot{}}
\makeatother

\usepackage{cite}      

\usepackage[pdftex]{graphicx}  

\usepackage{amsmath}   
\DeclareMathSizes{10}{10}{8}{7}   

\begin{document}

\title{Random Linear Network Coding For Time Division Duplexing: When To Stop Talking And Start Listening}

\author{\authorblockN{Daniel E. Lucani}
\authorblockA{RLE, MIT\\
Cambridge, Massachusetts, 02139\\
Email: dlucani@mit.edu}
\and
\authorblockN{Milica Stojanovic}
\authorblockA{Northeastern University\\
Boston, Massachusetts, 02118\\
Email: millitsa@mit.edu}
\and
\authorblockN{Muriel M\'edard}
\authorblockA{RLE, MIT\\
Cambridge, Massachusetts, 02139\\
Email: medard@mit.edu}
}


%


\maketitle
\begin{abstract}
A new random linear network coding scheme for reliable communications for time division duplexing channels is proposed. The setup assumes a packet erasure channel and that nodes cannot transmit and receive information simultaneously. The sender transmits coded data packets back-to-back before stopping to wait for the receiver to acknowledge (ACK) the number of degrees of freedom, if any, that are required to decode correctly the information. We provide an analysis of this problem to show that there is an optimal number of coded data packets, in terms of mean completion time, to be sent before stopping to listen. This number depends on the latency, probabilities of packet erasure and ACK erasure, and the number of degrees of freedom that the receiver requires to decode the data. This scheme is optimal in terms of the mean time to complete the transmission of a fixed number of data packets. We show that its performance is very close to that of a full duplex system, while transmitting a different number of coded packets can cause large degradation in performance, especially if latency is high. Also, we study the throughput performance of our scheme and compare it to existing half-duplex Go-back-N and Selective Repeat ARQ schemes. Numerical results, obtained for different latencies, show that our scheme has similar performance to the Selective Repeat in most cases and considerable performance gain when latency and packet error probability is high.
 
\end{abstract}


%
\IEEEpeerreviewmaketitle

\section{Introduction}

Network coding was introduced by Ahlswede \textit{et al} \cite{ahslwede00}. This concept is also known as coded packet networks. Network coding considers the nodes to have a set of functions that operate upon received or generated data packets. Today's networks would represent a subset of the coded packet networks, in which each node has two main functions: forwarding and replicating a packet. A classical network's task is to transport packets provided by the source nodes unmodified. In contrast, network coding considers information as an algebraic entity, on which one can operate.

Network coding research originally studied throughput performance without delay considerations for the transmitted information. 
The seminal work by Ahlswede \textit{et al} \cite{ahslwede00} considers a channel with no erasures and, therefore, no need for feedback. Work in \cite{li03} and \cite{medard03} showed that linear codes over a network are sufficient to implement any feasible multicast connection, again considering a channel with no erasures. Also, \cite{medard03} provides an algebraic framework for studying this subset of coded networks. In both of these cases, the nodes are considered to transmit a linear combination of the packets previously received. 
Work in \cite{ho06} presents the idea of using linear codes generated randomly in a network and shows that it achieves multicast capacity in a non-erasure channel. 

For networks with packet erasures, two approaches have been used. The first approach relies on rateless codes, i.e. transmitting coded data packets until the receiver sends an acknowledgement stating that all data packets have been decoded successfully. Reference \cite{lun08} studies random linear network coding in lossy networks showing that it can achieve packet-level capacity for both single unicast and single multicast connections in wireline and wireless networks. Reference \cite{maymounkov06} presents network codes that preserve the communication efficiency of a random linear code, while achieving better computational efficiency. Reference \cite{lun06wiopt} presented a random linear coding scheme for packet streams considering nodes with a fixed, finite memory, establishing a trade-off between memory usage and achievable rate. In terms of practical issues and implementation, work in \cite{chachulski07} presents MORE, a MAC-independent opportunistic protocol for wireless networks, and provides experimental results with some emphasis on the throughput gains provided by network coding. 

The work in \cite{eryilmaz06} and \cite{ahmed07} has studied delay performance gains and their scaling laws for network coding with and without channel side information, respectively. They focus on transmission of large files in a rateless fashion. In \cite{ghaderi07} the delay performance of network coding for a tree-based multicast problem is studied and compared to various Automatic Repeat reQuest (ARQ) and Forward Error Correcting (FEC) techniques. For network coding, it assumes reliable and instantaneous feedback to acknowledge a correct decoding of all data packets. 
Note that the focus of these references has been on either throughput or delay performance, usually considering minimal feedback. 

Finally, the work in \cite{jaykumar08} couples the benefit of network coding and ARQ by acknowledging degrees of freedom (dof), defined as linearly independent combinations of the data packets, instead of original data packets to show that queue size in a node follows degrees of freedom. 

The second approach uses block transmissions. Reference \cite{dana06} studies the problem in wireless networks and shows that linear codes achieve capacity in the network. In \cite{shrader07} a queueing model for random linear coding is presented, which codes data packets in a block-by-block fashion using acknowledgements to indicate successful transmission of each block. Interestingly, the coding block size depends on the number of packets available in the queue, up to some maximum block size.  

	
	We study channels in which time division duplexing is necessary, i.e. when a node can only transmit or receive, but not both at the same time. This problem has not been considered in any of the previous network coding references or, to the best of our knowledge, for network coding before our work. This type of channel is usually called half-duplex in the literature, but we use the term time division duplexing (TDD) to emphasize that the transmitter and receiver do not use the channel half of the time each or in any pre-determined fashion. Important examples of time division duplexing channels are infrared devices (IrDA), which have motivated many TDD ARQ schemes \cite{ozugur00} \cite{shah05}, and underwater acoustic communications \cite{milica05}. Other important applications may be found in channels with very high latency, e.g. in satellite \cite{sastry75}\cite{yu81}, and deep space \cite{Akyildiz04} communications.  

\begin{figure}[t]
\centering	
\includegraphics[height=1in,width=3.5in]{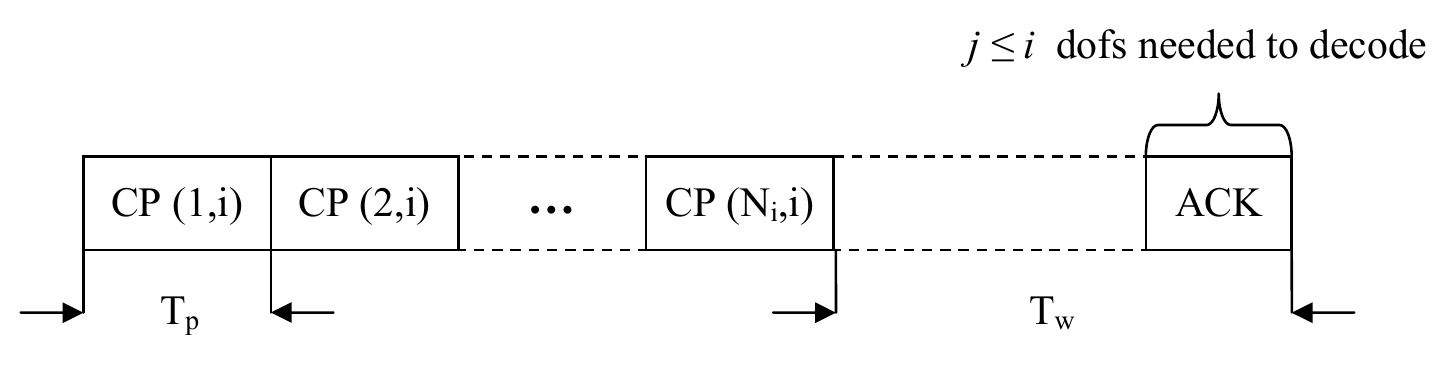}
\caption{Network coding TDD scheme}
\label{Protocol.tag}
\end{figure}

More specifically, we focus on the problem of transmitting $M$ data packets through a link using random linear network coding. 
We consider that the sender can transmit random linear coded packets back-to-back before stopping to wait for an acknowledgement (ACK) packet. This ACK packet conveys the remaining dofs required at the receiver to decode all $M$ data packets. We consider that
the number of coded packets $N_i$ to be transmitted before waiting for a new ACK packet depends on the number of dofs $i$ needed at the receiver, as indicated by the last ACK packet received successfully. If it is the first transmission, we consider that the required dofs is $M$. Figure~\ref{Protocol.tag} illustrates the communication process. The system transmits $N_{i}$ coded packets (CP),  and waits to receive an ACK packet that updates the value of $i$ to $j$, at which point it will transmit $N_{j}$ coded packets. The system will keep transmitting and stopping to update $i$, until $i$~=~0.  When $i$~=~0, the transmitter can start with $M$ new data packets, or simply stop. In Figure~\ref{Protocol.tag}, $CP(k,d)$ represents the $k$-th coded packet transmitted when we start transmission with $d$ dofs needed at the receiver to decode the information.

There is a natural trade off in the choice of the $N_i$'s. Every time the system stops to wait for an ACK, it incurs in an additional delay, which can be large in high latency channels. In general, the system requires at least one stop to get confirmation of complete transmission. However, we want to minimize the number of stops required to complete transmission of the $M$ packets. Note that if the $N_i$'s are too small given the channel conditions, the system will have to transmit more ACK packets to complete transmission of the block of $M$ data packets, which will cause a larger delay. On the other hand, if the $N_i$'s are too large, the receiver will have decoded the $M$ packets, for example, before the transmitter stops sending the first $N_M$ coded packets. Since the block of $M$ original packets is considered to be completely transmitted when the ACK requests no more dofs, the system causes unnecessary delay by transmitting too many coded packets by delaying transmission of the ACK by the receiver. In other words, the transmitter could have sent a smaller number of coded packets before stopping and still transmit the $M$ packets successfully.    

We show that there exists an optimal number of coded data packets to be transmitted back-to-back before stopping to wait for an ACK packet from the receiver, in terms of mean completion time, i.e. mean time to decode the $M$ original data packets at the receiver and get an ACK at the transmitter. In fact, the optimal number of coded data packets $N_i$ depends on the number of dofs $i$ that the receiver requires to decode the information, and also on the packet error probability and the latency, i.e. the number of bits in flight.  Thus, we show that there is an optimal time for stop transmitting coded packets and start listening to an ACK packet from the receiver.

	Thus, our objective is to minimize the expected time to complete transmission of a block, i.e. the delay in block transmissions, using feedback. This delay to decode a block is different from the usual packet delay measure. Since coding is carried out on blocks of packets, the delay to decode a block successfully determines the delay of each of the packets in that block. We also show that minimizing the expected time to complete transmission of a block of $M$ packets with a fixed packet size also maximizes the throughput performance. However, we show that a correct choice of $M$ and number of bits in the data packet can further improve throughput performance.

	Although both standard ARQ techniques and our scheme achieve reliability by detecting errors in received packets or packet erasures, and recover the information using a retransmission scheme, there are some important differences. First, we rely on transmission of coded packets, i.e. there is no need to specify a particular data packet to retransmit as in ARQ, but only a random linear combination. The ACK packet of our scheme thus differs from common ARQ techniques \cite{lin84} in that it does not give acknowledgement to particular data packets \cite{lin84}, but to degrees of freedom needed at the receiver to decode the $M$ original packets. Second, the number of coded packets transmitted in our scheme is not fixed by design of the algorithm, but chosen given channel characteristics and information in the ACK packet. In fact, the information in the ACK packet of our algorithm can be used to update an estimate of the probability of packet error and improve the overall performance. 

	
	We present an analysis and numerical results that show that transmitting the optimal number of coded data packets sent before stopping to listen for an ACK provides performance very close to that of a network coding scheme operating in a full duplex channel, in terms of mean time to complete transmission of all packets. This is the case even in high latency channels. Choosing a number different from the optimum can cause a large degradation in performance, especially if latency is high.


	Since random linear network coding is used, the results of this paper can be extended to the case of a network, in which each node performs a random linear combination of packets received from different nodes. In this extension, each node transmitting through a link, or, more generally, a hyperarc (using the terminology in \cite{lun06}) will have an optimal number of coded packets to transmit back-to-back before stopping to listen. 
	
	 The paper is organized as follows. In Section 2, we present the setup of the problem. In Section 3, we present the analysis of expected time to complete transmission of $M$ data packets and the optimization required to determine the number of coded packets to transmit before stopping. Also, two network coding comparison schemes are presented. In Section 4, the throughput performance is analyzed. In Section 5, numerical results are presented comparing our TDD optimal network coding with several other schemes in terms of the mean time to complete transmission. We also present results that compare throughput performance of our scheme to that of typical ARQ schemes. Conclusions are summarized in the Section 6.

\section{Random Network Coding for TDD channels}

\begin{figure}[t]
\centering	
\includegraphics[height=0.7in,width=3in]{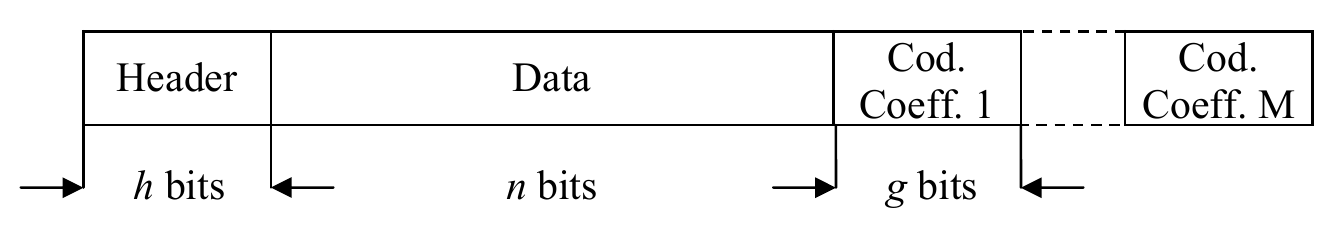}
\caption{Structure of coded data packet}
\label{PacketFrame.tag}
\end{figure}

	A sender in a link wants to transmit $M$ data packets at a given link data rate $R$. The channel is modeled as a packet erasure channel. Nodes can only transmit or receive, but not both at the same time. The sender uses random linear network coding \cite{ho06} to generate coded data packets. Each coded data packet contains a linear combination of the $M$ data packets of $n$ bits each, as well as the random encoding coefficients used in the linear combination. Each coefficient is represented by $g$ bits. For encoding over a field size $q$, we have that $g = \log_2 q$ bits. Also consider an information header of size $h$. Thus, the total number of bits per packet is $h + n + gM$. Figure \ref{PacketFrame.tag} shows the structure of each coded packet consider in our scheme. 

	The sender can transmit coded packets back-to-back before stopping to wait for the ACK packet. The ACK packet feeds back the number of degrees of freedom, that are still required to decode successfully the $M$ data packets. Since random linear coding is used, there is some probability of choosing encoding vectors that are all zero for one coded packet or encoding vectors that are linearly dependent on vectors of previously received packets. Thus, using arguments similar to \cite{eryilmaz06}, the expected number of successfully received packets before having $M$ linearly independent combinations, is
\begin{eqnarray}
\sum_{k =1}^M \frac{1}{(1 - {(1/q)}^k)} \leq M \frac{q}{q-1}
\end{eqnarray}

In the following analysis, we assume that the field size $q$ is large enough so that the expected number of successfully received packets at the receiver, in order to decode the original data packets, is approximately $M$. This is not a necessary assumption for our analysis. We could have included the probabilities of receiving linearly independent combinations into the transition probabilities. However, making this assumption simplifies the expressions and provides a good approximation for large enough $q$.

	We are interested in determining the optimal number of coded packets that should be sent back-to-back before waiting for an ACK packet from the receiver in order to minimize the time for successfully transmitting the $M$ data packets over the link.

\begin{figure}[t]
\centering	
\includegraphics[height=2.5in,width=2in]{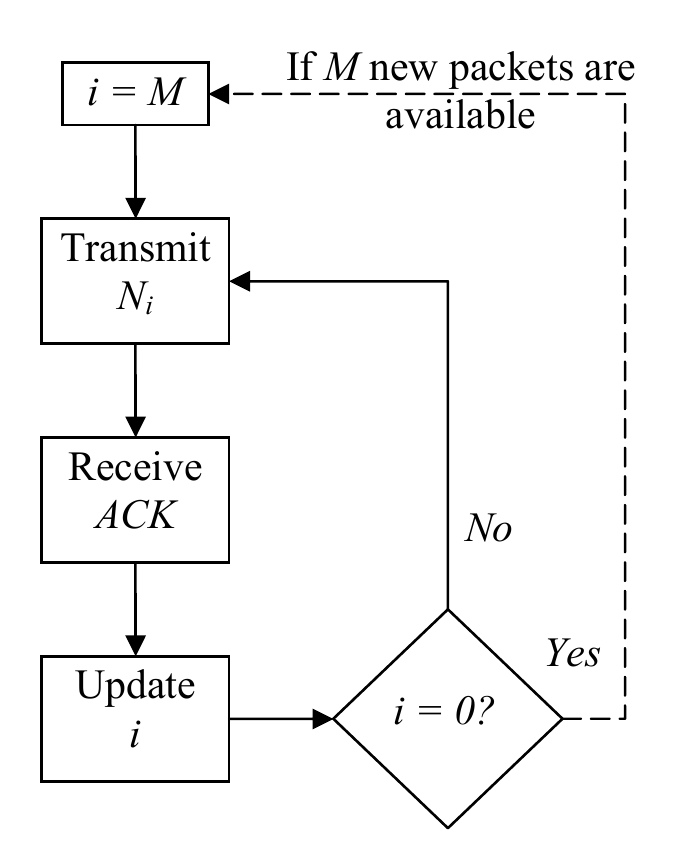}
\caption{Algorithm of network coding for time division duplexing channels. $i$ represents the remaining number of degrees of freedom to decode the packets, and $N_i$ the corresponding number of coded packets transmitted before stopping to listen for a new ACK. The ACK packet has the information to update $i$}
\label{algorithm.tag}
\end{figure}

\begin{figure}[h]
\centering	
\includegraphics[height=2in,width=3.5in]{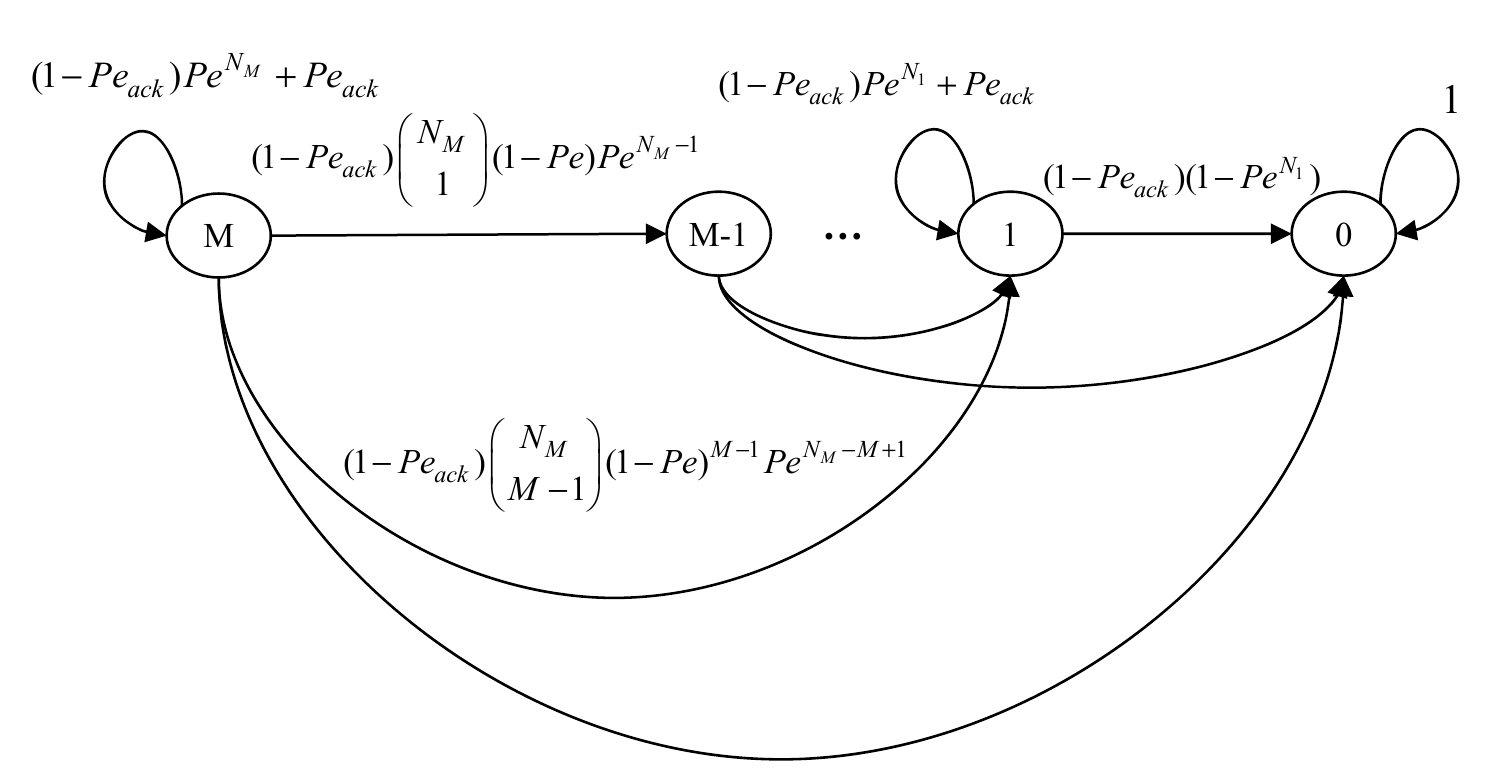}
\caption{Markov chain representation of the scheme. State $i$ represents that the receiver requires $i$ more successfully received coded packets to decode the information}
\label{MarkovChain.tag}
\end{figure}    	
	
	Note that if $M$ packets are in the queue, at least $M$ degrees of freedom have to be sent in the initial transmission, i.e. $N_M \geq M$ coded packets. We are interested not only in the number of dof that are required at the first transmission, but also at subsequent stages. Transmission begins with $M$ information packets, which are encoded into $N_M$ random linear coded packets and transmitted. If all $M$ packets are decoded successfully, the process is completed. Otherwise, the ACK informs the transmitter how many are missing, say $i$. The transmitter then sends $N_i$ coded packets, and so on, until all $M$ packets have been decoded successfully. We are interested in the optimal number $N_{i}$ of coded packets to be transmitted back-to-back in the next transmission to complete the remaining $i$ dof's. 
Figure~\ref{algorithm.tag} shows the communication process as a system transmits $N_{M}$ coded packets initially and awaits reception of an ACK packet that updates the value of $i$, at which point it will transmit $N_{i}$ coded packets. The system will keep transmitting and stopping to update $i$, until $i$~=~0.  When $i$~=~0, the transmitter can start with $M$ new data packets or simply stop. In Figure~\ref{Protocol.tag}, $CP(k,d)$ represents the $k$-th coded packet transmitted when we start transmission with $d$ dofs needed at the receiver to decode the information.

	The process can be modelled as a Markov Chain (Figure~\ref{MarkovChain.tag}). The states are defined as the number of dof's required at the receiver to decode successfully the $M$ packets. Thus, these states range from $M$ to 0. This is a Markov Chain with $M$ transient states and one recurrent state (state 0). Let us define $N_{i}$ as the number of coded packets that are sent when $i$ dof's are required at the receiver in order to decode the information. Note that the time spent in each state depends on the state itself, because $N_{i} \neq N_{j}, \forall i \neq j$ in general.       
	
	The transition probabilities from state $i$ to state $j$ ($P_{i\rightarrow j}$) have the following expression for $0<j< i$ and $N_i \geq i$:

\begin{eqnarray}
P_{i\rightarrow j} = (1-Pe_{ack})\binom{N_{i}}{i - j} {(1-Pe)}^{i-j} {Pe}^{N_{i} -i + j}
\end{eqnarray}
where $Pe$ and $Pe_{ack}$ represents the erasure probability of a coded packet and of an ACK packet, respectively.

More generally, the transition probability can be defined for any value of $N_i \geq 1$ as follows:

\begin{eqnarray}
P_{i\rightarrow j} = (1-Pe_{ack}) f(i,j) {(1-Pe)}^{i-j} {Pe}^{N_{i} -i + j}
\end{eqnarray}		
where 

\begin{align}
f(i,j) =
\begin{cases}
 \binom{N_{i}}{i - j} & \text{if $N_{i} \geq i$,}
\\
0 & \text{otherwise}
\end{cases}
\end{align}

For $j = i$ the expression for the transition probability reduces to:

\begin{eqnarray}
P_{i\rightarrow i} = (1-Pe_{ack}){Pe}^{N_{i}} + Pe_{ack}
\end{eqnarray}		

\section{Expected Time for completing transmission} \label{SectionExpectedTime.tag}

	The expected time for completing the transmission of the $M$ data packets constitutes the expected time of absorption, i.e. the time to reach state 0 for the first time, given that the initial state is $M$. This can be expressed in terms of the expected time for completing the transmission given that the Markov Chain is in state is $i$, $T_{i}$ , $\forall i = 0, 1, .. M - 1$. Let us denote the transmission time of a coded packet as $T_{p}$, and the waiting time to receive an ACK packet as $T_{w}$. For our scheme, $T_{p} = \frac{h +n+gM}{R}$ and $T_{w} = T_{rt} + T_{ack}$, where $T_{ack} = n_{ack}/R$, $n_{ack}$ is the number of bits in the ACK packet, $R$ is the link data rate, and $T_{rt}$ is the round trip time. Note that $T_{0} = 0$. Then, for $i>1$:
	
\begin{eqnarray}
T_{i} &=& \frac{ N_{i} T_{p} + T_{w}}{(1-Pe_{ack})(1 - Pe^{N_{i}})}\\ 
&+& \frac{  {(1 - Pe)}^i {Pe}^{N_{i} - i} \sum _{j = 1} ^{i-1} f(i,j) {\left( \frac{Pe}{1 - Pe} \right)}^j T_{j}}{1 - Pe^{N_{i}}}
\end{eqnarray} 
 
	For example, for $i$~=~1 we have that:
\begin{eqnarray}
T_{1} = \frac{\left( N_{1} T_{p} + T_{w} \right)}{(1-Pe_{ack})(1 - Pe^{N_{1}})} 
\end{eqnarray}  
As it can be seen, the expected time for each state $i$ depends on all the expected times for the previous states. Because of the Markov property, we can optimize the values of all $N_{i}$'s in a recursive fashion, i.e. starting by $N_{1}$, then $N_{2}$ and so on, until $N_{M}$, in order to minimize the expected transmission time. We do so in the following subsection.      
 
\subsection{Minimizing Expected Time for Completing Transmission} 

	Our objective is to minimize the value of the expected transmission time $T_{M}$. Under the assumption that $N_{i} \geq i$, we have:

\begin{eqnarray}
&&\min_{N_{M},..,N_{1}} T_{M} = \notag \\&& \min_{N_{M},..,N_{1}}\scriptstyle   \frac{ N_{M} T_{p} + T_{w}}{(1-Pe_{ack})(1 - Pe^{N_{M}})} \normalsize \\ &&+ \scriptstyle \frac{ {(1 - Pe)}^M {Pe}^{N_{M} - M} \sum _{j = 1} ^{M-1} \binom{N_{M}}{M - j} {\left( \frac{Pe}{1 - Pe} \right)}^j T_{j}}{1 - Pe^{N_{M}}} \normalsize \notag \\
&&= \min_{N_{M}} \scriptstyle \frac{  N_{M} T_{p} + T_{w}}{(1-Pe_{ack})(1 - Pe^{N_{M}})} \normalsize \\&&+ \scriptstyle  \frac{  {(1 - Pe)}^M {Pe}^{N_{M} - M} \sum _{j = 1} ^{M-1} \binom{N_{M}}{M - j} {\left( \frac{Pe}{1 - Pe} \right)}^j  \min_{N_{j},..,N_{1}}  T_{j}}{1 - Pe^{N_{M}}} \normalsize \notag
\end{eqnarray} 
Without this assumption, we have that

\begin{eqnarray}
&&\min_{N_{M},..,N_{1}} T_{M} =\notag\\&& \min_{N_{M},..,N_{1}} \scriptstyle  \frac{N_{M} T_{p} + T_{w}}{(1-Pe_{ack})(1 - Pe^{N_{M}})} \normalsize \\&& + \scriptstyle  \frac{ {(1 - Pe)}^M {Pe}^{N_{M} - M} \sum _{j = 1} ^{M-1} f(M,j) {\left( \frac{Pe}{1 - Pe} \right)}^j T_{j}}{1 - Pe^{N_{M}}} \normalsize \notag\\
&&= \min_{N_{M}} \scriptstyle   \frac{  N_{M} T_{p} + T_{w} }{(1-Pe_{ack})(1 - Pe^{N_{M}})} \normalsize \\&&+ \scriptstyle \frac{ {(1 - Pe)}^M {Pe}^{N_{M} - M} \sum _{j = 1} ^{M-1} f(M,j) {\left( \frac{Pe}{1 - Pe} \right)}^j  \min_{N_{j},..,N_{1}}  T_{j}}{1 - Pe^{N_{M}}} \normalsize \notag
\end{eqnarray} 
Hence, regardless of the assumption on $N_i$, the problem of minimizing $T_{M}$ in terms of the variables $N_{M},..,N_{1}$ can be solved iteratively. First, we compute $\min_{N_{1}}  T_{1}$, then use this results in the computation of $\min_{N_{2},N_{1}}  T_{2}$, and so on.

	One approach to computing the optimal values of $N_{i}$ is to ignore the constraint to integer values and take the derivative of $T_{i}$ with respect to $N_{i}$ and look for the value that sets it equal to zero. For our particular problem, this approach leads to solutions without a closed form, i.e. expressed as an implicit function. For $M = 1$, the optimal value of $N_1$ can be expressed using a known implicit function (Lambert function), and it is given by 
\begin{eqnarray}
N_{1}^* = \frac{ 1 + W \left( - \exp{\left( -1 + \frac{\ln(Pe) T_{w}}{T_{p}}  \right)} \right)    }{\ln {Pe} }  - \frac{T_{w}}{T_{p}}
\end{eqnarray}	
where $W(\cdot)$ is the Lambert W function \cite{Chapeau}. The positive values are found for the branch $W_{-1}$, as denoted in reference \cite{Chapeau}. 

	The case of $M = 1$ can be thought as an optimized version of the uncoded Stop-and-Wait ARQ, which is similar to the idea presented in \cite{sastry75}. Instead of transmitting one packet and waiting for the ACK, our analysis suggests that there is an optimal number of back-to-back repetitions of the same data packet that should be transmitted before stopping to listen for an ACK packet. 

 Instead of using the previous approach, we perform a search for the optimal values $N_i, \forall i \in \{1,...M\}$, using  integer values. Thus, the optimal $N_i$'s can be computed numerically for given $Pe$, $Pe_{ack}$, $T_{w}$ and $T_{p}$. In particular, the search method for the optimal value can be made much simpler by exploiting the recursive characteristic of the problem, i.e. instead of making a $M$-dimensional search, we can perform $M$ one-dimensional searches. Finally, these $N_i$'s do not need to be computed in real time. They can be pre-computed and store in the receiver as look-up tables. This procedure reduces the computational load on the nodes at the time of transmission.

\subsection{Comparison Scheme 1: Fixed Maximum Window}	

	Let us consider the same setting, i.e. a fixed number of packets $M$ that have to be transmitted to the receiver, but with a fixed, pre-determined maximal number of coded packets to be transmitted before stopping to listen. We define this maximal value of coded packets as $ \omega $. If the number of degrees of freedom $i$ required at the receiver to decode the information is  $i \geq \omega$, the transmitter will transmit $\omega$ degrees of freedom. If $i < \omega$, the transmitter will transmit $i$ degrees of freedom. 
	
	The model for the Markov Chain is derived from the previous case, by setting $N_{i} = \omega, \forall i \geq \omega$ and $N_{i} = i, \forall i< \omega$. For $i \geq \omega$, we have that:

\begin{eqnarray}
&&T_{i} = \frac{ \omega T_{p} + T_{w}}{(1-Pe_{ack})(1 - Pe^{\omega})}\\&& + \frac{  \sum _{j = 1} ^{\omega} \binom{\omega}{j} {\left( {Pe}^{\omega - j}{(1 - Pe)}^j \right)} T_{i-j}}{1 - Pe^{\omega}} 
\end{eqnarray} 		
and for $i < \omega$:
\begin{eqnarray}
&&T_{i} = \frac{ i T_{p} + T_{w}}{(1-Pe_{ack})(1 - Pe^{i})}\\&& + \frac{ \sum _{j = 1} ^{i} \binom{i}{j} {\left( {Pe}^{i - j}{(1 - Pe)}^j \right)} T_{i-j}}{1 - Pe^{i}} 
\end{eqnarray} 		

\subsection{Comparison Scheme 2: Optimal Full Duplex ARQ }	

	This scheme assumes that nodes are capable of receiving and transmitting information simultaneously, and in that sense it is optimal in light of minimal delay. The sender transmits coded packets back-to-back until an ACK packet for correct decoding of all information ($M$ information packets) has been received. This scheme can be modeled as a Markov Chain where, as before, the states represent the number of dofs received. The time spent in each state is the same ($T_{p}$). Once the $M$ packets have been decoded, i.e. $M$ dofs have been received, the receiver transmits ACK packets back-to-back, each of duration $T_{ack}$. One ACK should suffice but this procedure minimizes the effect of a lost ACK packet.
	
	The mean time to complete the transmission and get and ACK is:
\begin{eqnarray}
E[T] = T_{rt} + \frac{M T_{p}}{1 - Pe} + \frac{T_{ack}}{1-Pe_{ack}}	
\end{eqnarray} 		
where $T$ is the time to complete transmission of $M$ packets.

\section{Throughput}
		
		The mean throughput is defined as $E[\frac{Mn}{T}]$, where $T$ is the time to complete transmission of $M$ packets. For $Mn$ deterministic we have $Mn E[\frac{1}{T}]$. For the case of $M=1$, i.e. the extended version of the Stop-and-Wait ARQ scheme, we can provide a simple expression for the mean throughput in terms of the transition probabilities $P_{1\rightarrow 1}$ and $P_{1\rightarrow 0}$, 
\begin{eqnarray}
E[\frac{1}{T}] &= \frac{P_{1\rightarrow 0}}{P_{1\rightarrow 1}} \sum_{k = 1}^{\infty } \frac{{P_{1\rightarrow 1}}^k}{k \left(T_{p} + T_{w}   \right)} \\&=  \frac{P_{1\rightarrow 0}}{P_{1\rightarrow 1}\left(T_{p} + T_{w}   \right)}  \sum_{k = 1}^{\infty } \frac{{(1 - P_{1\rightarrow 0})}^k}{k}\\& = -\frac{P_{1\rightarrow 0}}{P_{1\rightarrow 1}\left(T_{p} + T_{w}   \right)} \ln (P_{1\rightarrow 0})
\end{eqnarray}

We have used the Mercator series since $|1-P_{1\rightarrow 0}| < 1$ for all cases of interest. However, for $M>1$ this expressions are complicated. 
	Thus, we define our measure of throughput $\eta$ as the ratio between number of data bits transmitted ($N$) and the time it takes to transmit them. For the case of a block-by-block transmission, as described in Section~\ref{SectionExpectedTime.tag},
\begin{eqnarray}\label{throughput.tag}
\eta =  \frac{ M n }{ T_{M} } 
\end{eqnarray} 		
where $T_{M}$ is the expected time of completion defined previously.     

Note that the expected throughput and $\eta$ are not equal. For the case of $M = 1$, note that $E[\frac{Mn}{T}] = \eta \frac{\ln (1/P_{1\rightarrow 0})}{P_{1\rightarrow 1}}$. More generally, using Jensen's inequality,  $Mn E[\frac{1}{T}] \geq \frac{Mn}{T_M}$ for $T>0$. Therefore, $\eta$ constitutes a lower bound to the mean throughput in our scheme. 
Another reason to consider this measure is to compare our network coding scheme with typical ARQ schemes that do not rely on coded packets since the analysis for most ARQ schemes is performed using $\eta$.
		
	Note that if $M$ and $n$ are fixed, $\eta$ is maximized as $T_M$ is minimized. Thus, by minimizing the mean time to complete transmitting of a block of $M$ data packets with $n$ bits each, we are also maximizing $\eta$ for those values. However, we show that the maximal $\eta$ should be obtained using $M$ and $n$ as arguments in our optimization.
	
	This is important for systems in which the data is streamed. In this case, searching for the optimal values of $M$ and $n$, in terms of $\eta$, provides a way to optimally divide data into blocks of $M$ packets with $n$ bits each before starting communication using our scheme.

\subsection{Optimal Packet size and packets per block}
	We have discussed throughput with a pre-determined choice of the number of data bits $n$ and the number of data packets $M$ in each block. 
	However, expression \ref{throughput.tag} implies that the throughput $\eta$ depends on both $n$ and $M$. Hence, it is possible to choose these parameters so as to maximize the throughput. We can approach this problem is several ways. 
The first approach is to look for the optimal $n$ while keeping $M$ fixed:
\begin{eqnarray}
\eta_{opt}(M) = \arg \max_n \left \{   \max_{N_M,...,N_1} \eta   \right \}
\end{eqnarray} 		

The second approach is to look for the optimal $M$ while keeping $n$ fixed:
\begin{eqnarray}
\eta_{opt}(n) = \arg \max_M \left \{   \max_{N_M,...,N_1} \eta   \right \}
\end{eqnarray} 		

	More generally, we could consider the case in which both parameters are variable and we are interested in maximizing $\eta$:
\begin{eqnarray}
\eta_{opt} = \arg \max_{n,M} \left \{   \max_{N_M,...,N_1} \eta   \right \}
\end{eqnarray} 		
	
\section{Numerical Examples}
	
	This section provides numerical examples that compare the performance of the different network coding schemes we have discussed so far in TDD channels. The comparison is carried out in terms of the mean time to complete a transmission of $M$ data packets through TDD channel under different block error probabilities. We also present results in terms of the measure of throughput $\eta$ to illustrate its dependence on the values of $M$ and $n$ for varying channel characteristics  (erasure probabilities). We use the case of satellite communications as an example of high latency channels. 

\begin{figure}[t]
\centering	
\includegraphics[height=3.5in,width=3.5in]{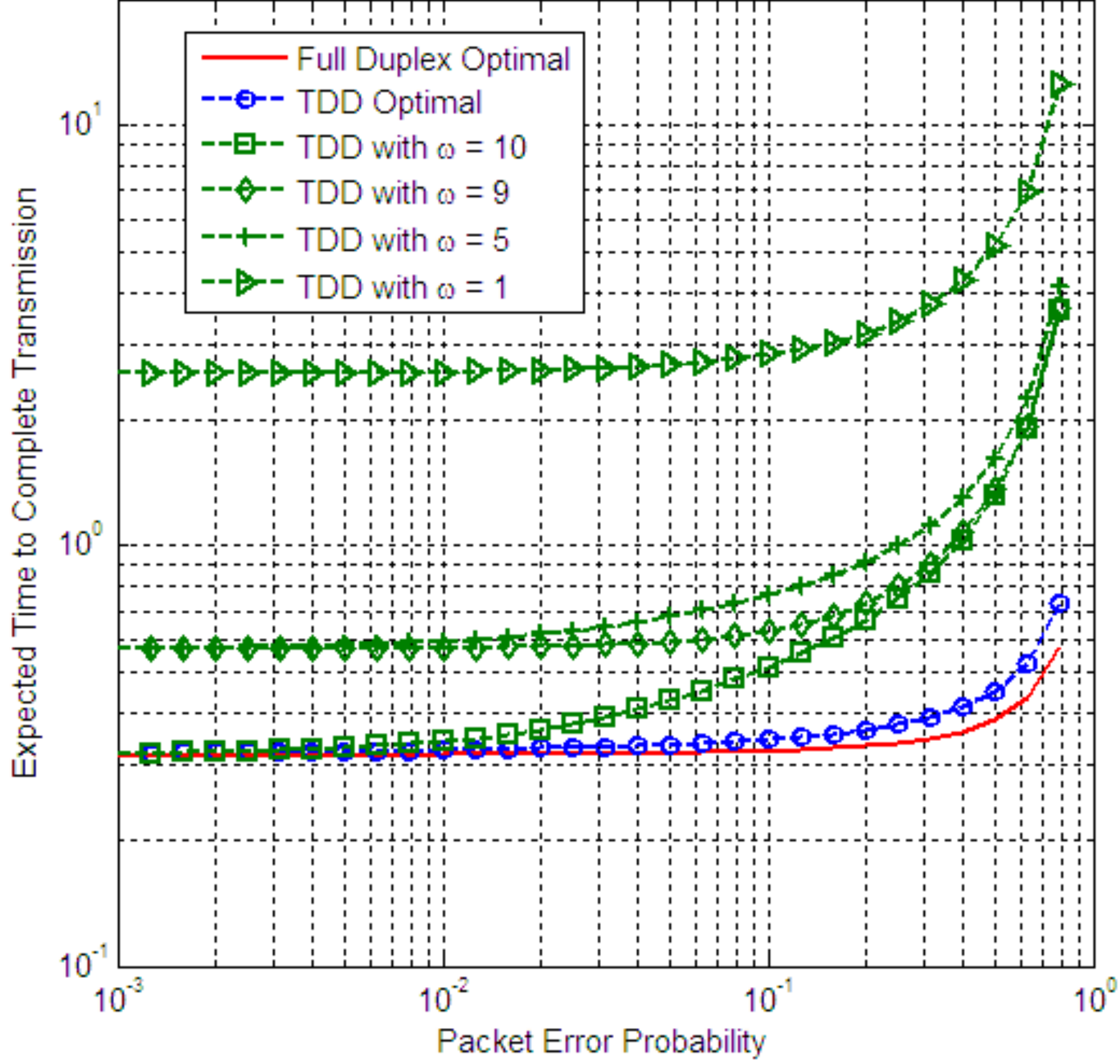}
\caption{Expected time for transmitting $M$ data packets successfully versus $Pe$ in a satellite example. The parameters used are $M = 10$, $T_{rt} = 250$~ms, data rate $1.5$~Mbps, $n_{ack} = 100$~bits, $n = 10000$~bits, $g = 100$~bits, $h = 80$~bits, $Pe_{ack} = 0.001$}
\label{ExpectedTimeLongPropagation.tag}
\end{figure}

Figure ~\ref{ExpectedTimeLongPropagation.tag} studies the expected time to complete transmission of $M = 10$ data packets of size $n = 10000$ bits, with different packet error probabilities in a GEO satellite link with a propagation delay of 125~ms.
 We assume a link with parameters specified in the figure. Note that our network coding scheme (TDD optimal) and the network coding full duplex optimal scheme have similar performance over a wide range of block error probabilities. In fact, for the worst case ($Pe = 0.8$) presented in this figure, our scheme has an expected time of completion only 29~\% above the full duplex scheme. This is surprising considering that the transmitter in the full duplex scheme sends coded packets non-stop until an ACK packet is received. The explanation for this behavior is that our scheme is sending enough coded packets, given the channel conditions, so that the number of stops to listen (which are very costly) is minimized. Thus, our scheme can have similar performance to that of full duplex optimal scheme, in the sense of expected time to completion. Most importantly, our scheme is very likely to have a much better performance in terms of energy consumption due to the long periods in which the transmitter stops to listen for the ACK packets.

\begin{figure}[t]
\centering	
\includegraphics[height=3.5in,width=3.5in]{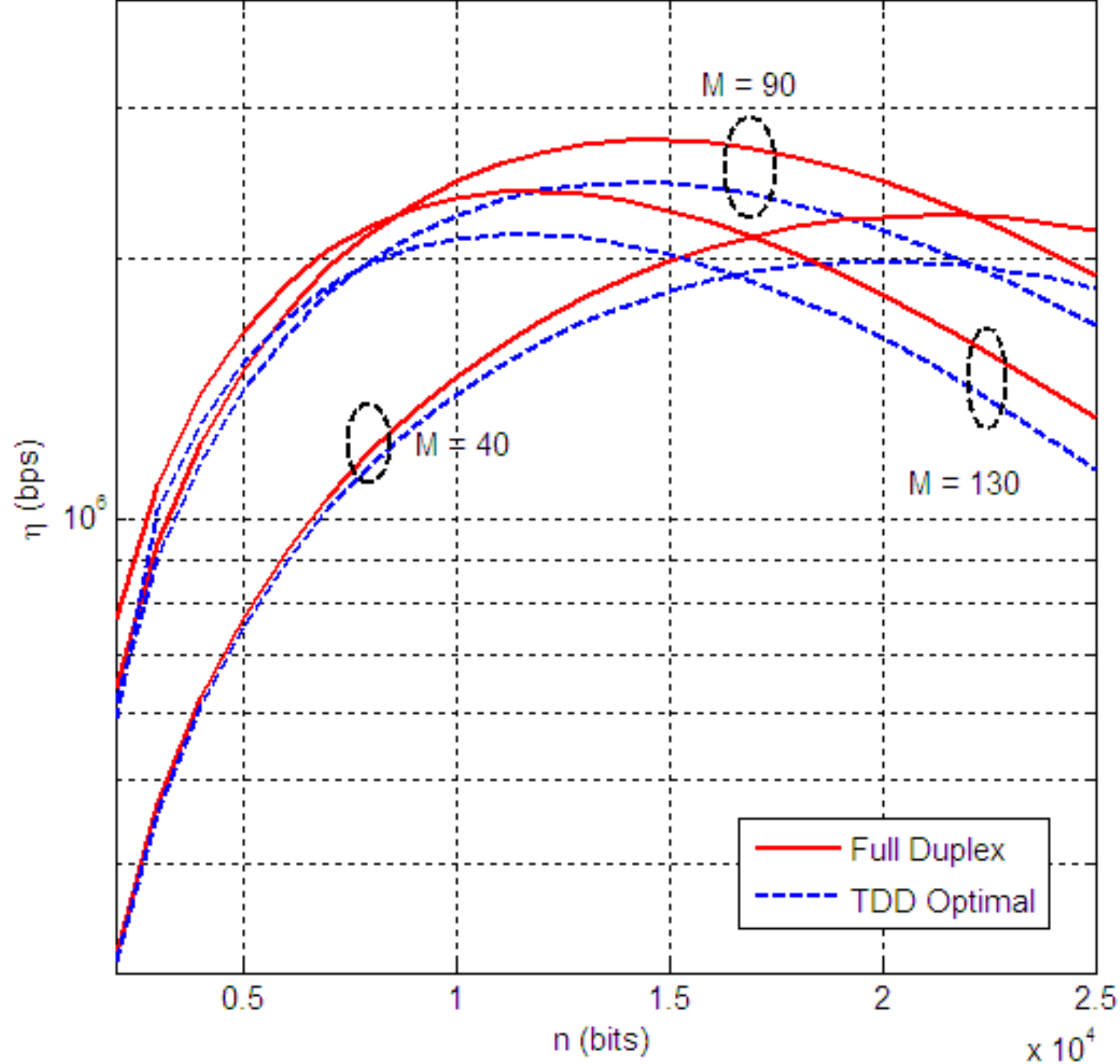}
\caption{Throughput measure $\eta$ versus the number of bits $n$ in a data packet for a symmetrical channel, for different values of $M$ with parameters $g~=~100$~bits, $n_{ack}~=~100$~bits, $h = 80$ bits, data rate $100$~Mbps, $T_{rt} = 250$~ms, $Pe_{bit} = 0.0001$}
\label{ChangingMSymmetricChannel.tag}
\end{figure}    

\begin{figure}[h]
\centering	
\includegraphics[height=3.5in,width=3.5in]{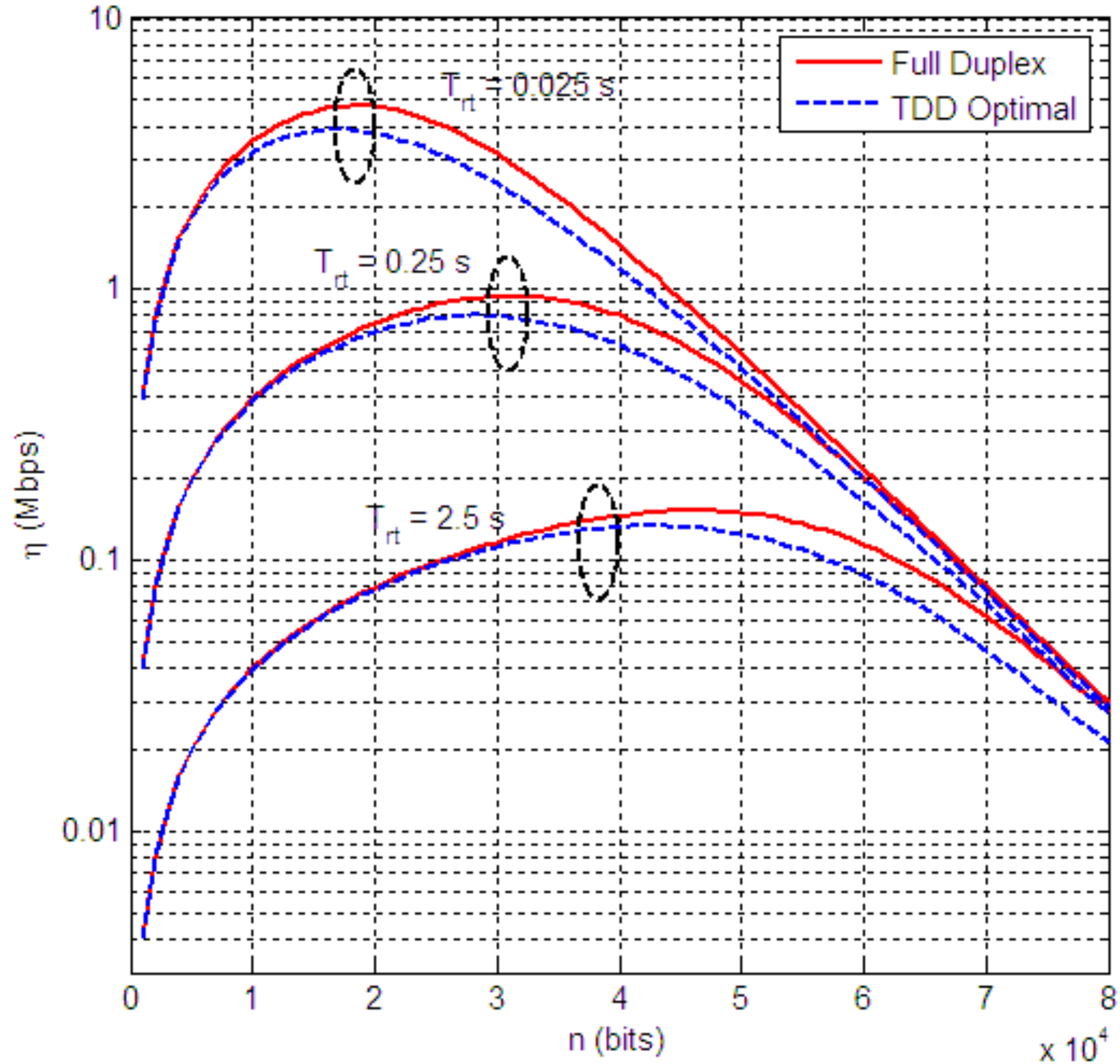}
\caption{Throughput $\eta$ versus $n$ in a symmetrical channel considering different values of round-trip time $T_{rt}$ with parameters $g~=~100$~bits, $n_{ack}~=~100$~bits, $h = 80$~bits, data rate $1.5$~Mbps, $M=10$, $Pe_{bit} = 0.0001$}
\label{ChangingPropagationTimeSymmetricChannel.tag}
\end{figure}

\begin{figure}[h]
\centering	
\includegraphics[height=3.5in,width=3.5in]{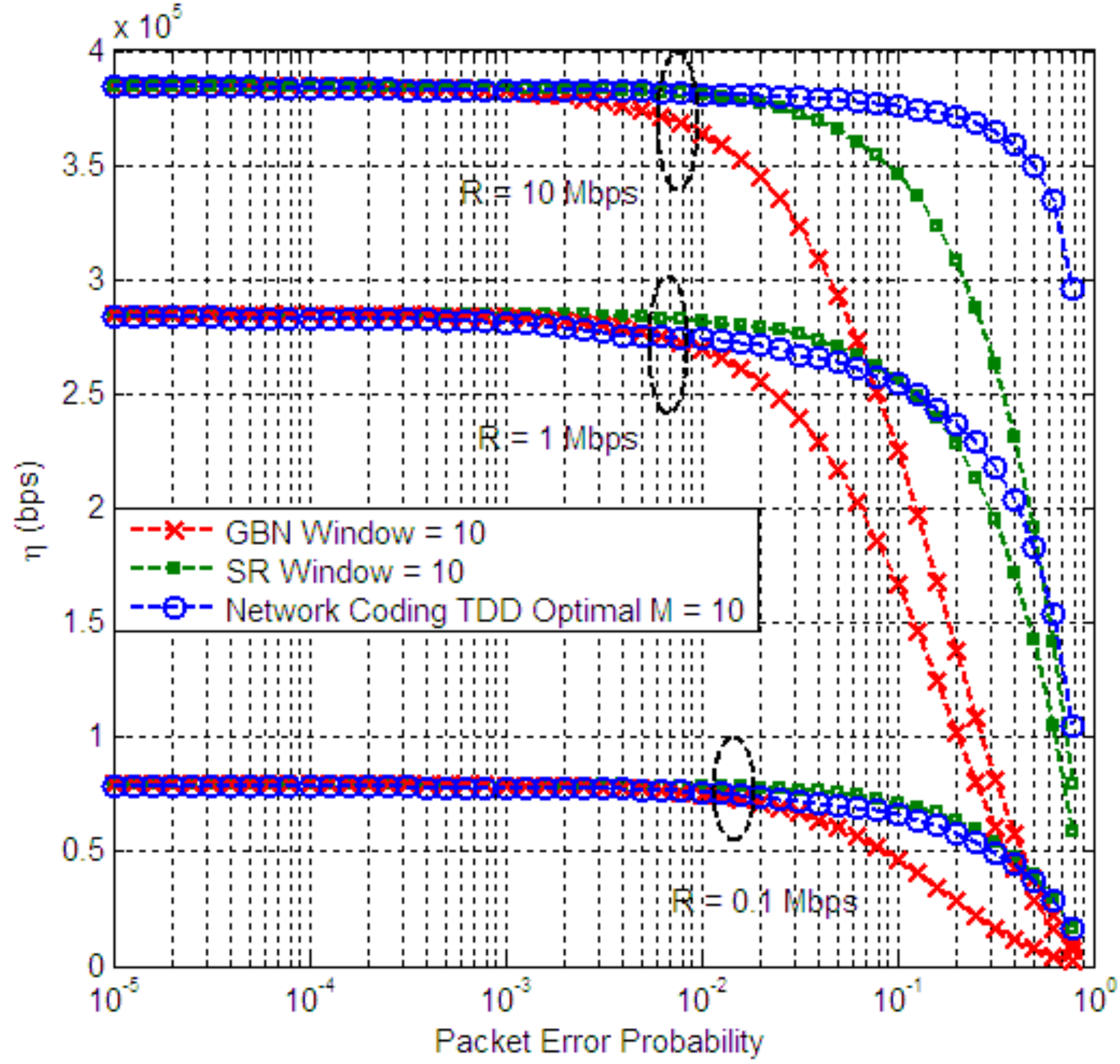}
\caption{$\eta$ versus data packet error probability with two TDD non-network coding schemes (Go-Back-N and Selective Repeat) and our optimal TDD network coding scheme, with different $R$. We used as parameters $g~=~20$~bits, $n_{ack}~=~100$~bits, $n~=~10000$~bits, $h = 80$~bits, $T_{rt} =$~0.25~ms}
\label{halfduplexdifferentschemes.tag}
\end{figure}

\begin{figure}[h]
\centering	
\includegraphics[height=3.5in,width=3.5in]{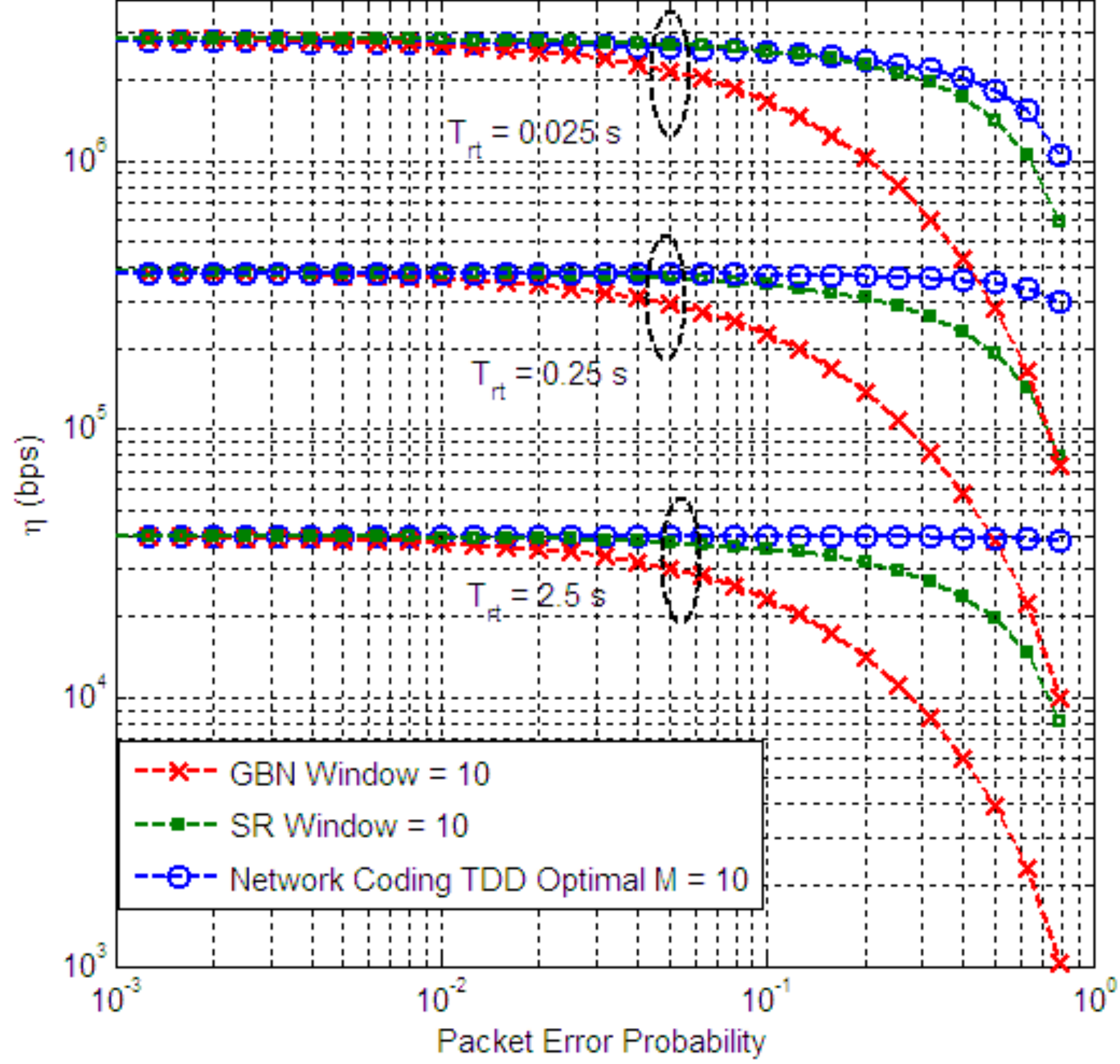}
\caption{$\eta$ versus data packet error probability with two TDD non-network coding schemes (Go-Back-N and Selective Repeat) and our optimal TDD network coding scheme, with different $T_{rt}$ values. We used as parameters $g~=~20$~bits, $n_{ack}~=~100$~bits, $n~=~10000$~bits, $h = 80$~bits, $R=$~10~Mbps}
\label{halfduplexdifferentschemesn_10000_DR_10Mbps_Trt.tag}
\end{figure}

   
Figure ~\ref{ExpectedTimeLongPropagation.tag} also shows the performance of the comparison scheme 1 presented in Section III. Note that when $\omega = 10$, i.e. the transmitter sends at most 10 coded packets before stopping to listen, the performance is comparable to our optimal scheme when the block error probability is low. This fact confirms that for low block error probabilities the optimal choice of coded packets to transmit when $i$ dof are required at the receiver ($N_i$) is simply $i$. In other words, if $M=10$ and the block error probability is low, the first transmission contains 10 coded packets. Note that using $\omega = 9$ already suffers from a considerable degradation in performance even for low $Pe$ because the transmitter cannot transmit the minimum number of coded packets ($M$) necessary to decode the information after the first transmission, and so it must transmit at least one more coded packet after the first ACK. Note that the performance of $\omega = 5$ and $\omega = 9$ is similar for low block error probability because both of them require at least two stops to listen for ACK packets in order to relay all the information, and it is the stopping time that affects delay the most on a high latency channel. For the case of $\omega>10$ we would see a degradation for low $Pe$, with respect to optimum, because more packets than necessary are transmitted. 

Finally, note that for the worst data error probability in Figure ~\ref{ExpectedTimeLongPropagation.tag}, all fixed schemes (TDD with fixed $\omega$) take at least 5 times more time to complete transmission than the network coding full duplex optimal scheme. The case of $\omega = 1$ can be interpreted as the performance of the Stop-and-Wait ARQ scheme under the same channel conditions, which is considerably worse than the other schemes.

	Let us turn our attention now to the problem of maximizing the parameter $\eta$, i.e. our mean throughput lower bound. Recall that for this setting we are streaming data which is subdivided into blocks that are transmitted them using our scheme. Considering again a satellite link, given a fixed bit error probability ($Pe_{bit} = 0.0001$) let us study the problem of computing the optimal number of bits $n$ per packet given some value of $M$. In these examples, for the case of a symmetric channel with independent bits $Pe = 1 - (1-Pe_{bit})^{h + n + gM}$ and $Pe_{ack} = 1 - (1-Pe_{bit})^{n_{ack}}$.

Figure~\ref{ChangingMSymmetricChannel.tag} illustrates the values of $\eta$ in Mbps given different choices of $M$ and $n$. First, note that for each value of $M$ there exists an optimal value of $n$. Thus, an arbitrary choice of $n$ can produce a considerable degradation in performance in terms of throughput. Secondly, there is a $(M,n)$ pair that maximizes the value of $\eta$. Finally, the performance of the full duplex network coding and our TDD optimal scheme is comparable for different values of $n$ and $M$.

Figure~\ref{ChangingPropagationTimeSymmetricChannel.tag} shows $\eta$ in Mbps when we change the round-trip time $T_{rt}$. As expected, a lower $T_{rt}$ allows more throughput in TDD. Again, we observe that our TDD optimal scheme has comparable performance to the full duplex scheme. 

Let us compare the performance of our optimal TDD network coding scheme with respect to typical TDD ARQ schemes: Go-back-N (GBN) and Selective Repeat (SR). For this comparison, we use the $\eta$ factor for the half-duplex version's of these schemes. Reference \cite{ozugur00} studied both of these cases and proposed the utilization factor. In our notation, the equivalent $\eta$'s are given by $\eta_{GBN}$ and $\eta_{SR}$ for GBN and SR, respectively:
\begin{eqnarray}
\eta_{GBN} =   \frac{ n(1-Pe)\left(1 - {(1-Pe)}^W \right)}{ (WT_{p} + T_{w})Pe}
\end{eqnarray} 		
and
\begin{eqnarray}
\eta_{SR} =   \frac{ Wn(1-Pe) }{ WT_p + T_w} 
\end{eqnarray} 
where $W$ is the window size.

Figure~\ref{halfduplexdifferentschemes.tag} shows $\eta$ for the satellite communications setting with a fixed packet size of $n = 10000$ bits, $n_{ack} = 100$ bits, $T_{rt} = 250$ ms, $Pe_{ACK} = 0$ for all schemes, a window size of $W=10$ for the ARQ schemes, and $g = 20$ bits and $M = 10$ for our network coding scheme. We use different data rates to illustrate different latency scenarios, where higher data rate is related to higher latency. Note that the performance of our scheme is the same as both GBN and SR at low data packet error probability, which is expected because the window size $W$ is equal to the block size of our scheme $M$ and we expect very few errors. Our scheme has a slightly lower $\eta$ for low $Pe$ because each coded data packet includes $gM$ additional bits that carry the random encoding vectors. This effect is less evident as latency increases. In general, our scheme has better performance than GBN.  

Figure~\ref{halfduplexdifferentschemes.tag} shows that for low latency (0.1~Mbps) $\eta$ of our scheme is very close to that of the SR ARQ scheme for all values of $Pe$, and better than the GBN scheme for high $Pe$. These results are surprising, because our scheme constitutes a block-by-block transmission scheme which will not start transmission of a new set of $M$ data packets until the previous ones have been received and acknowledged. Note also that, as latency increases, our scheme shows much better performance than the SR scheme for high $Pe$. The case of 10~Mbps and $Pe = 0.8$ shows that $\eta$ of our scheme is more than three (3) times greater than that of SR.    

Figure~\ref{halfduplexdifferentschemesn_10000_DR_10Mbps_Trt.tag} shows $\eta$ for a fixed data rate of $10$~Mbps and different $T_{rt}$. We use a fixed packet size of $n = 10000$ bits, $n_{ack} = 100$ bits, $Pe_{ACK} = 0$ for all schemes, a window size of $W=10$ for the ARQ schemes, and $g = 20$ bits and $M = 10$ for our network coding scheme. Note that the overhead of transmitting $M$ coefficients of $g$ bits per coded packet is only 2\%. Thus, this effect cannot be appreciated in the figures. Again, the performance of our scheme is the same as both GBN and SR at low data packet error probability. Since the data rate is kept fixed, at higher $T_{rt}$ we get higher latency. The throughput performance is similar to that observed in Figure~\ref{halfduplexdifferentschemes.tag} if we carry our comparison in terms of latency.     

	Another advantage of our scheme with respect to SR ARQ is that our scheme relies on transmitting successfully one block of $M$ data packets before transmitting a new one. In fact, our scheme minimizes the delay of every block. In contrast, the SR ARQ does not provide any guarantee of delay for any data packet, e.g. the first packet of a file to be transmitted could be the last one to be successfully received. In this sense, our comparison is not completely fair, as it favors the standard schemes. Nonetheless, our scheme is providing similar or better performance than SR but guaranteeing low transmission delays in individual data packets.
	
\section{Conclusion}

	This paper proposes a new random linear network coding scheme for reliable communications for time division duplexing channels. This scheme optimizes the mean time to complete transmission of a number of data packets by determining the number of coded data packet that the sender has to transmit back-to-back before stopping to wait for the receiver to acknowledge how many degrees of freedom, if any, are required to decode correctly the information. 

	The optimal number of coded data packets, in terms of mean completion, to be sent back-to-back depends of the latency, probabilities of erasure of the coded packet and the ACK, and the number of degrees of freedom that the receiver requires to decode the data. While there is no closed form solution for the optimal number of packets, we can perform a search of the optimal values. In particular, the search method for the optimal value is simple by exploiting the recursive characteristic of the problem, i.e. instead of making a $M$-dimensional search, we perform $M$ one-dimensional searches. Finally, these values do not need to be computed in real time. They can be pre-computed and stored in the receiver as look-up tables. This procedure makes the computational load on the nodes to be negligible at the time of determining the optimal transmission time.
	
	We present means of analysis and numerical results to show that transmitting the optimal number of coded packets before stopping to listen for an ACK is very close to the performance of a full duplex system, while choosing a different number can cause considerable degradation in performance, especially if latency and packet error probability are high. Notably, our scheme also shows good potential to improve energy consumption. 

	In terms of throughput performance, we compare our scheme to the standard half-duplex Go-back-N and Selective Repeat ARQ schemes. Numerical evaluation for different latency shows that our scheme has similar performance to the Selective Repeat in most cases, and considerable performance gain when latencies and packet error probability are high. Numerical results also show that our scheme is superior to Go-back-N when error probability is high for different latency. 
	 
	Future research will consider the problem of 1) energy consumption and associated optimization for this scheme, and 2)  extension of the principles proposed (which were analyzed for one link in a network) to the general problem of wireless networks.
   

\section*{Acknowledgment}
This work was supported in part by the National Science Foundation under grants No. 0520075, 0427502, and CNS-0627021, by ONR MURI Grant No. N00014-07-1-0738, and subcontract \# 060786 issued by BAE Systems National Security
Solutions, Inc. and supported by the Defense Advanced Research Projects
Agency (DARPA) and the Space and Naval Warfare System Center (SPAWARSYSCEN),
San Diego under Contract No. N66001-06-C-2020 (CBMANET).

\end{document}